\documentclass[11pt,a4paper]{article}

\usepackage{graphicx}
\usepackage{caption}
\usepackage{authblk}
\usepackage[left=2.5cm,right=2.5cm,top=1.5cm,bottom=1.5cm,includeheadfoot]{geometry}
\usepackage{bm}
\usepackage{amsmath}

\begin{document}

\title{High-resolution two-dimensional electronic spectroscopy reveals homogeneous line profiles in isolated nanoparticles}%
% High-resolution two-dimensional electronic spectroscopy of phthalocyanine molecules in cryogenic nano-environments
\author{Ulrich Bangert}
\author{Frank Stienkemeier}
\author{Lukas Bruder\thanks{lukas.bruder@physik.uni-freiburg.de}}%

\affil{Institute of Physics, University of Freiburg, Hermann-Herder-Str. 3, 79104 Freiburg, Germany.}

\maketitle

\begin{abstract}
Doped clusters in the gas phase provide nanoconfined model systems for the study of system-bath interactions. To gain insight into interaction mechanisms between chromophores and their environment, the ensemble inhomogeneity has to be lifted and the homogeneous line profile must be accessed. However, such measurements are very challenging at the low particle densities and low signal levels in cluster beam experiments. 
Here, we dope cryogenic rare-gas clusters with phthalocyanine molecules and apply action-detected two-dimensional electronic spectroscopy to gain insight into the local molecule-cluster environment for solid and superfluid cluster species. 
The high-resolution homogeneous linewidth analysis provides a benchmark for the theoretical modelling of binding configurations and shows a promising route for high-resolution molecular two-dimensional spectroscopy. 
\end{abstract}

\maketitle

\section{Introduction}
The control and spectroscopic study of nanoconfined systems is of high relevance in many fields of research, ranging from quantum technologies to light harvesting applications, catalysis and spectroscopy\,\cite{li_all-optical_2003, garnett_nanowire_2011, zhang_surface-plasmon-driven_2018, aeschlimann_perfect_2015}. 
To understand the structure-function relationship on the nanoscale, individual structural configurations have to be resolved with high resolution and the influence of parasitic environmental perturbations are minimized. 
These challenges are best met in studies of isolated nano-objects in the gas phase. 

In cluster-isolation spectroscopy rare-gas clusters serve as nanoconfined matrix to isolate individual spectroscopic probes in the gas phase\,\cite{toennies_superfluid_2004, stienkemeier_spectroscopy_2006,kupper_spectroscopy_2007, mauracher_cold_2018, verma_infrared_2019}. 
Helium (He) clusters form superfluid nanodroplets which efficiently cool dopant species to sub-Kelvin internal temperatures inside a homogeneous solvent, providing ideal conditions for high-resolution spectroscopy\,\cite{stienkemeier_electronic_2001, choi_infrared_2006}, e.g. of isolated molecules\,\cite{toennies_superfluid_2004}, radicals\,\cite{kupper_spectroscopy_2007}, aggregates\,\cite{roden_vibronic_2011} and complexes\,\cite{gutberlet_aggregation-induced_2009, higgins_photoinduced_1996, nauta_nonequilibrium_1999}.
More recently, solid rare-gas clusters where employed as a nanoconfined cryogenic substrate facilitating the formation of molecular networks with tunable interaction strength on the cluster surface\,\cite{muller_cooperative_2015}. 
This approach has revealed cooperative molecular mechanisms such as superradiance and singlet fission with spectral resolution not achievable in thin film and bulk experiments\,\cite{muller_cooperative_2015, izadnia_singlet_2017}. 
While the cluster-isolation technique was so far predominantly used for high-resolution spectroscopy of embedded species, the heterogeneous cluster systems may also serve as models for the study of structural configurations in isolated nanoconfined systems. 

Insight into system-bath interactions, e.g. the coupling to the bath modes, the co-existence and configuration of different binding sites and their dynamic rearrangement are generally gained from the absorption line profiles\,\cite{friedrich_photochemical_1984}. 
Thereby, the key challenge is the extraction of the homogeneous line profile from the inhomogeneously broadened ensemble response. 
However, most methods for homogeneous linewidth retrieval\,\cite{friedrich_photochemical_1984} are not compatible with the low particle densities of gas-phase cluster samples. 
Hole burning, as an exception, has been applied to doped He nanodroplets, however did not resolve the homogeneous linewidth\,\cite{hartmann_hole-burning_2001}. 
This technique faces the difficulty of adapting the laser parameters to the time and frequency scales of the target system, reducing the approach mainly to photochemical hole burning where photoreactive chromophores are probed on quasi infinite time scales\,\cite{friedrich_photochemical_1984}. 
These challenges may explain why in heterogeneous cluster samples the homogeneous linewidth has not been determined, so far. 

Recently, two-dimensional electronic spectroscopy (2DES) has been established, which is an ultrafast nonlinear spectroscopy technique enabling the disentanglement of homogeneous and inhomogeneous broadenings while automatically adapting to the time-frequency scale of the probed ensemble\,\cite{jonas_two-dimensional_2003}. 
2DES and 2D infrared spectroscopy\,\cite{hochstrasser_two-dimensional_2007} have proven as very useful in condensed-phase systems to extract line shape information where other methods are not applicable or do not provide the required time-frequency resolution\,\cite{cowan_ultrafast_2005, moca_two-dimensional_2015, ryu_relations_2021, liu_toward_2021, aeschlimann_coherent_2011}. 
Due to technical challenges, the method's potential in gas phase experiments is hardly explored\,\cite{bruder_coherent_2018,roeding_coherent_2018, bruder_coherent_2019}. 

Here, we apply the method to a cluster-isolated chromophore in the gas phase, which enables us to resolve the homogeneous absorption profile of the system. 
As chromophore we chose free-base phthalocyanine (H$_2$Pc), which belongs to a class of aromatic molecules of high relevance in optoelectronics, nonlinear opical materials and photobiology\cite{claessens_phthalocyanines:_2008, mckeown_phthalocyanine_1998, sorokin_phthalocyanine_2013, hoppe_organic_2004, amerongen_photosynthetic_2000}. 
Our results provide insight into the molecule-surface binding configurations with a resolution far beyond the accuracy and resolution of current density functional theory approaches. 
This offers a new perspective for resolving the role of local configurations in photo-chemical reactions and opens a route for ultrafast multidimensional spectroscopy studies of isolated molecular systems with unprecedented spectral resolution.

\section{Results}
\subsection{Experimental scheme}
\begin{figure}
\centering
  \includegraphics{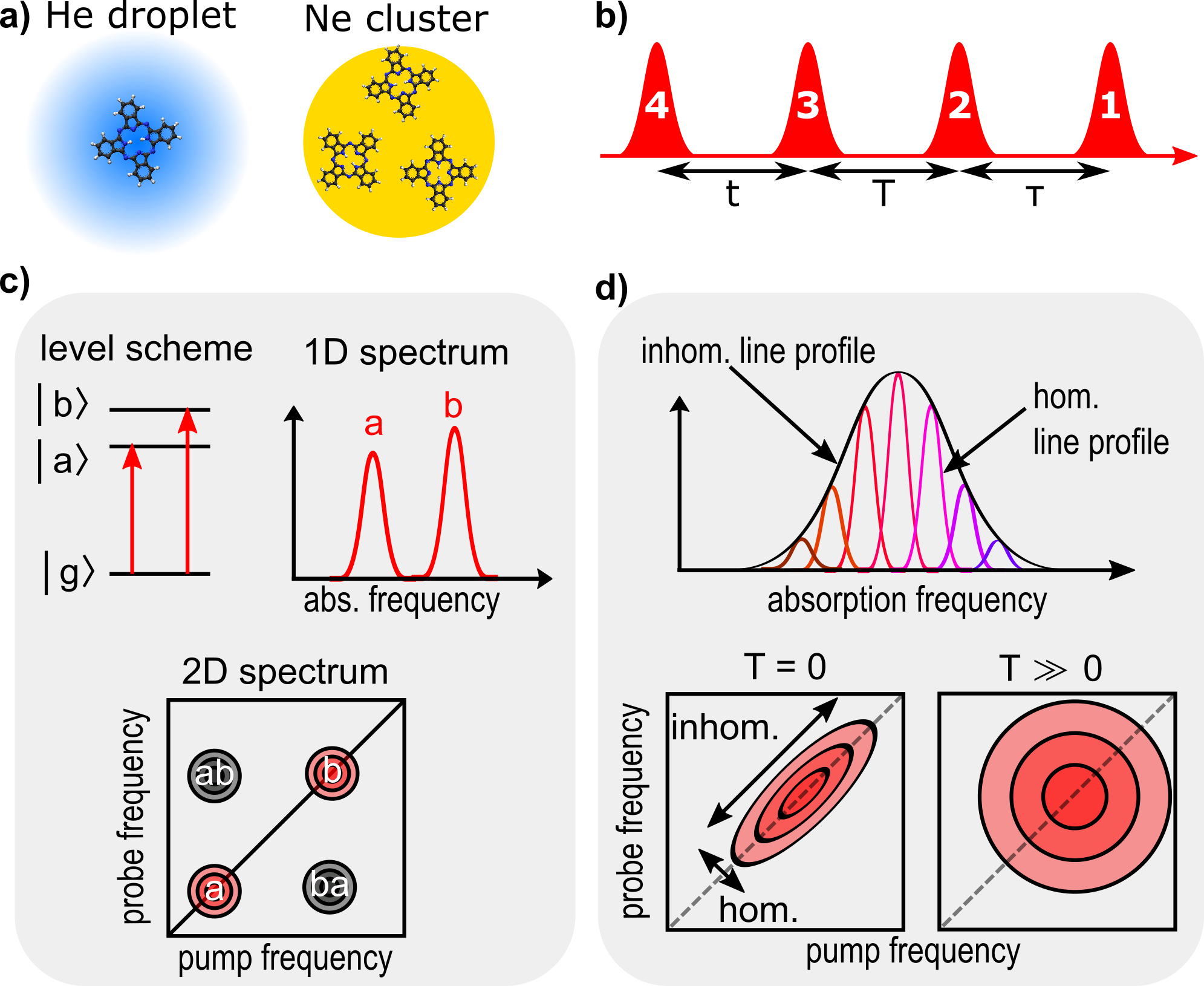}
  \caption{Experimental scheme. 
  (a) Studied sample types: single H$_2$Pc molecule dissolved in a He droplet (blue) and $\leq 3$ molecules attached to the surface of a solid Ne cluster (yellow). 
  (b) Optical femtosecond pulse train used in the 2DES experiment. 
  (c,d) Relevant signal contributions in the 2D spectra. 
  In (c): the linear excitation spectrum is mapped onto the diagonal of the 2D spectrum  (peaks labeled a,b), while nonlinear couplings between the quantum states (e.g. through a common ground state) lead to off-diagonal peaks (labels ab, ba). 
  In (d): Depending on the local ensemble configuration, the homogeneous absorption of the molecules is frequency-shifted, leading to an inhomogeneously broadened lineshape. 
  In the 2D spectra, the inhomogeneous (homogeneous) lineprofile is aligned along the diagonal (anti-diagonal), and their correlation can be studied as a function of the evolution time $T$.}
  \label{fig1}
\end{figure}

Fig.\,\ref{fig1} summarizes the experimental scheme. 
A supersonic beam of rare-gas clusters is generated in a molecular beam apparatus and is doped with H$_2$Pc molecules (details in Methods section). 
In our study, we compare the molecular response of a single chromophore dissolved inside a superfluid He droplet with the response of 2--3 chromophores attached to the surface of a solid Ne cluster (Fig.\,\ref{fig1}a). 
At the low equilibrium temperatures of the nano-systems (He droplet: 0.37\,K\,\cite{hartmann_rotationally_1995}, Ne cluster: $\approx 10$\,K\,\cite{farges_structure_1981}), only the lowest vibrational state of the dopant molecule is thermally occupied. 
Nearest-neighbour interactions between the chromophores are well-suppressed in both cluster-isolation experiments and no spectral signatures (line shifts/splittings) of inter-molecular couplings are observed despite the high spectral resolution of the experiment. 

So far, 2DES is mainly performed in the condensed phase and the desired nonlinear signals are separated from the background by non-collinear four-wave mixing geometries (coherent-detected 2DES)\,\cite{fuller_experimental_2015}. 
Conversely, for the dilute cluster beam samples a collinear beam geometry is needed combined with the detection of an action signal (action-detected 2DES)\,\cite{bruder_coherent_2019}. 
In the latter, the sample is excited with a sequence of four femtosecond laser pulses (Fig.\,\ref{fig1}b) and the fourth-order light-matter response is deduced from the detected fluorescence. 
The pulse delays $\tau, t$ are interferometrically scanned to track the free polarization decay of the sample induced by pulse 1 and 3, respectively. 
Accordingly, a Fourier transform with respect to $\tau, t$ yields 2D frequency-spectra, which directly correlate the excitation (x-axis) and detection (y-axis) frequencies, while the time delay $T$ determines the time evolution of the correlation spectra (Fig.\,\ref{fig1}c,d)\,\cite{jonas_two-dimensional_2003}. 

The ultralow optical density of doped cluster beam samples ($OD\sim 10^{-11}$\,\cite{bruder_coherent_2018}), requires a highly sensitive 2DES apparatus\,\cite{bruder_coherent_2019}, which adapts the phase modulation technique developed by Marcus and co-workers\,\cite{tekavec_fluorescence-detected_2007}. 
This method implements carrier-envelope-phase modulation of the optical pulses on a shot-to-shot basis at a high laser repetition rate (200\,kHz). 
The phase modulation leads to characteristic beat notes in the fluorescence yield based on which the linear and nonlinear signal contributions are efficiently separated and amplified using lock-in detection. 
Thereby, the lock-in detection suppresses phase noise in the interferometric measurement and greatly enhances the detection sensitivity (Details in Method section). 
These properties enabled several studies of highly dilute samples\,\cite{bruder_phase-modulated_2015, bruder_coherent_2018, bruder_delocalized_2019} and quantum interference measurements at extremely short wavelengths\,\cite{wituschek_tracking_2020, wituschek_phase_2020}. 

The relevant information content of the 2D spectra is schematically shown in Fig.\,\ref{fig1}\,c,\,d. 
Peaks on the diagonal position of the 2D spectra closely resemble the linear excitation spectrum of the system, whereas off-diagonal peaks directly disclose nonlinear couplings, which are normally obscured in linear absorption measurements (Fig.\,\ref{fig1}c)\,\cite{lewis_probing_2012}. 
In addition, 2D lineshapes provide insight into ensemble inhomogeneities and their dynamic evolution (Fig.\,\ref{fig1}d). 
With the coherent multipulse excitation scheme, nonlinear rephasing and non-rephasing signals are recorded which leads to a separation of the inhomogeneous(homogeneous) lineshape along the diagonal(anti-diagonal) projection of the 2D spectra, accordingly. 
At time $T=0\,$fs this effect leads to an elongated peakshape indicating a strong correlation between the absorption and detection frequency. 
For $T>0$ statistical fluctuations of the local environment lead to a loss of the correlation resulting in increasingly symmetric peak shapes (termed spectral diffusion). 
This concept is known from 2D photon echo spectroscopy as performed in coherence-detected 2DES\,\cite{jonas_two-dimensional_2003, lewis_probing_2012}. 
The equivalence between coherence and action-detected 2DES in terms of lineshape information was shown in Ref.\,\cite{tan_theory_2008}. 
We note that coherence and action-detected 2DES may differ in the detection of excited state absorption and multiple quantum coherence signals\,\cite{maly_signatures_2018, maly_coherently_2020} which are, however, not relevant in the current study.
\begin{figure*}
\centering
 \includegraphics[width = \textwidth]{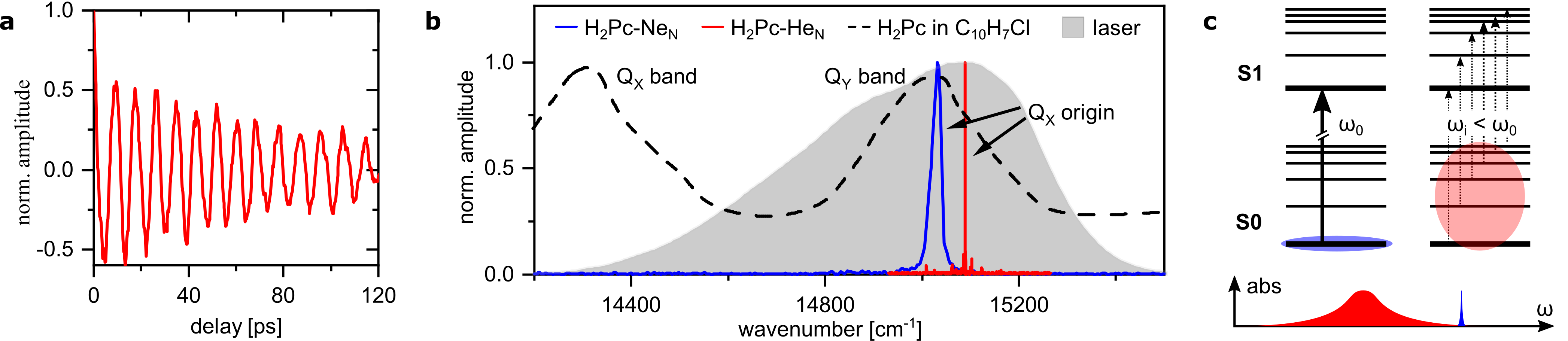}
 \caption{
    1D coherence scan of cluster-isolated H$_2$Pc. 
 (a) Long-lived electronic coherence of H$_2$Pc--He$_N$. The coherence oscillation is frequency-shifted to a lower domain due to rotating frame detection. In return, the frequency axis in the Fourier spectra in (b) are up-shifted by the same amount to recover the absolute transition frequencies of the system (see Methods section). 
 (b) Steady-state absorption spectrum of H$_2$Pc solvated in 1-chloronaphthalene at room temperature ($22\pm 1^\circ$C)\cite{jerwin_solvent_1984} (dashed black) compared to the Fourier spectra of the 1D coherence scans for the H$_2$Pc--He$_N$ (red) and H$_2$Pc--Ne$_N$ (blue), used femtosecond laser spectrum (grey). 
 The shown H$_2$Pc--He$_N$ linewidth is limited by the experimental resolution (see Method section). 
  (c) Excitation scheme of cold (blue) and hot (red) H$_2$Pc. 
 The shaded area illustrates the thermal population of vibrational states. 
 For hot molecules many vibronic absorption lines exist leading to a broadened red shifted absorption spectrum relative to the one obtained for cold molecules, as schematically shown in the absorption spectrum at the bottom.
 }
 \label{fig2}
\end{figure*}

\subsection{1D coherence scans}
So far, the majority of 2DES studies have been performed in the liquid and solid phase\,\cite{bruder_coherent_2019}. 
To relate our approach to experiments in the condensed phase, we perform 1D coherence scans (using pulse 1 and 2 only) of H$_2$Pc attached to Ne clusters (denoted H$_2$Pc--Ne$_N$) and He nanodroplets (denoted H$_2$Pc--He$_N$) and compare the Fourier spectrum to the linear absorption spectrum of H$_2$Pc in an organic solvent (Fig.\,\ref{fig2}). 
Fig.\,\ref{fig2}a shows the 1D coherence scan of the S$_1 \leftarrow$ S$_0$ transition in H$_2$Pc--He$_N$, featuring a clean, long-lived oscillation of the electronic coherence in the time domain. 
While in the condensed phase, electronic coherences are strongly perturbed and decay typically within less than 100\,fs\,\cite{cao_quantum_2020}, they persist for more than 100\,ps in H$_2$Pc--He$_N$ (Fig.\,\ref{fig2}a) and $\approx 3\,$ps in H$_2$Pc--Ne$_N$ (not shown), implying a weak coupling of the cluster environments to the molecule's electronic degrees of freedom. 
We note, that the full coherence decay in H$_2$Pc--He$_N$ extends beyond the experimental observation window. 
The observation of a single oscillating frequency in Fig.\,\ref{fig2}a suggests, that the molecule is initially prepared in a single state and the laser field drives a single vibronic transition. 
This offers ideal conditions for coherent control applications as well as optical trapping and cooling of molecules\,\cite{brumer_principles_2003, di_rosa_laser-cooling_2004}. 

A Fourier transform of the 1D coherence scans yield the linear absorption spectra of both systems (Fig.\,\ref{fig2}b). 
For reference we show the H$_2$Pc absorption spectrum in a 1-chloronaphthalene solution. 
Comparing the molecular response in the three different environments, a clear trend can be observed. 
An increasing line broadening and red shift of the absorption band occurs when going from H$_2$Pc--He$_N$ to H$_2$Pc--Ne$_N$ and H$_2$Pc in 1-chloronaphthalene.  
This can be explained by the reduced perturbation and lower temperature in the cluster-isolated molecular samples (cf. (Fig.\,\ref{fig2}c). 
A zoom on the Q$_x$ absorption band of the cluster-isolated molecules is shown in the Supporting Information along with a comparison to high-resolution steady-state laser excitation spectra. 

In the H$_2$Pc--He$_N$ spectrum several sharp peaks at much lower amplitude are observable clearly separated from the Q$_x$ absorption. 
These lines are assigned to complexes of H$_2$Pc and clusters of H$_2$O, N$_2$ or O$_2$ as confirmed by a comparison to previous steady-state excitation spectra of H$_2$Pc\,\cite{fischer_unpublished_2020} and other dopant molecules\,\cite{slenczka_inhomogeneous_2001, fischer_heterogeneous_2019}. 
While the superfluid, homogeneous He solvent favors the formation of heterostructures, complex formation on the surface of solid Ne clusters is less likely, explaining the absence of respective spectral signatures in this system.

\subsection{2D spectra}
\begin{figure}
\centering
  \includegraphics{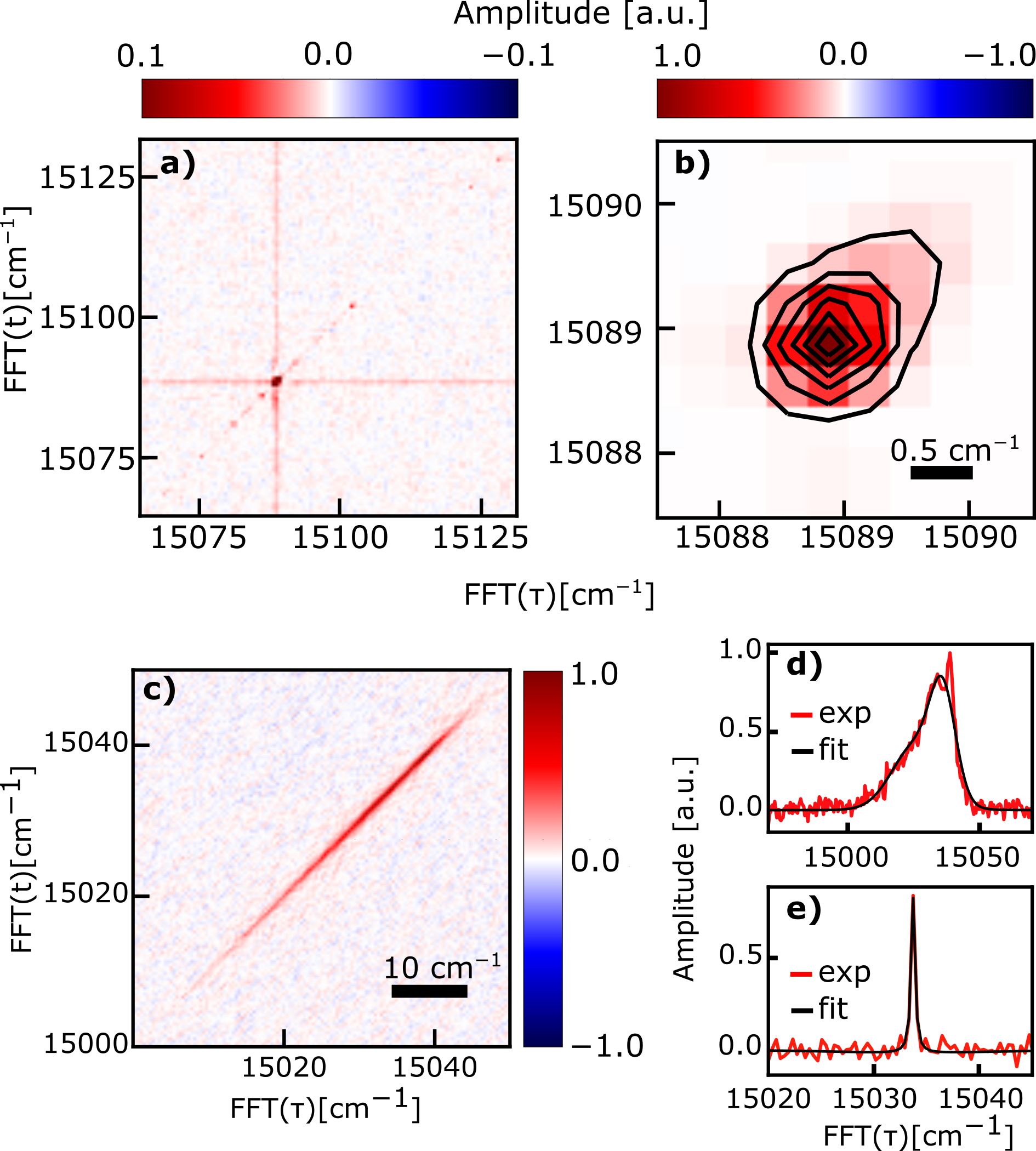}
  \caption{2D spectra at T=1\,ps of H$_2$Pc dissolved in superfluid helium nanodroplets (a, b) and adsorbed on the surface of solid neon clusters (c). 
  In (a) the color scale is saturated by a factor of 10, (b) shows an unsaturated zoom of the same data with contour lines at 14.2\% of the maximum amplitude. 
  (d, e) show the inhomogeneous(homogeneous) line profiles obtained by a cut of (c) along the diagonal axis (for d) and anti-diagonal axis at 15034cm$^{-1}$ (for e). 
  Black solid lines show the projections of the fit function obtained from a 2D lineshape fit.}
  \label{fig3}
\end{figure}

Having compared the linear spectra to studies in solution, we turn now to the nonlinear 2DES experiments. 
Fig.\,\ref{fig3}a shows the 2D spectrum of H$_2$Pc--He$_N$ featuring several sharp, highly-resolved diagonal peaks and no discernible off-diagonal resonances. 
Note, that the color scale is saturated by a factor of 10 to enhance the weaker spectral features. 
The strong diagonal peak at 15088.9\,cm$^{-1}$ marks the S$^0_1 \leftarrow$ S$^0_0$ transition in H$_2$Pc, i.e. the zero-phonon line (ZPL). 
The horizontal and vertical lines intersecting with the ZPL show the tails of the ZPL line profile, which are augmented by the saturated color scale. 
The weaker diagonal peaks are assigned to the heteromolecular complexes discussed above whose spectral positions coincide exactly with the weak features observed in the 2D spectra. 
Possible contributions from sample scattering can be excluded in action-detected 2DES\,\cite{maly_fluorescence-detected_2021}. 
The absence of cross-peaks confirms that these spectral features correspond to individual molecular species which do not couple among each other (cf. Fig.\,\ref{fig1}c). 
Since in the cryogenic, superfluid matrix clustering of dopant molecules is unavoidable, the absence of cross-peaks imply that the molecular complexes each occupy a separate nanodroplet.  

Fig.\,\ref{fig3}b shows a zoom on the 2D spectrum (unsaturated color scale), revealing the ZPL in more detail. 
The excitation and detection-frequency resolution in this 2D spectrum is 0.65\,cm$^{-1}$ (for resolution limit see Methods Section) which is, to the best of our knowledge, much higher than in any previously reported molecular 2DES study. 
Even higher resolution may be achieved with frequency comb-based 2DES\,\cite{lomsadze_frequency_2017}. 
However so far, this approach has been only demonstrated for thermal atomic vapors and not for cold molecules.
A spectral shoulder extending to the high-frequency side is observable which reveals the contribution of a C$_{13}$-isotope of H$_2$Pc located at 15089.5\,cm$^{-1}$ (details in the Supplementary Information). 
Taking this contribution into account, the ZPL reveals a homogeneous line profile (spherically symmetric 2D peak shape) within the experimental resolution. 
This confirms the high homogeneity of the quantum fluid environment also predicted by recent density-functional theory calculations\,\cite{ancilotto_density_2017}. 

As already deduced from the 1D absorption spectrum, the 2DES data confirms that the system behaves in good approximation like a two-level system with only a single optical transition and no coupling to other states (absence of cross-peaks). 
This is remarkable for organic molecules and is a consequence of the narrow Franck-Condon window for the S$_1 \leftarrow$ S$_0$ transition in combination with the low internal temperature of the molecule. 

For H$_2$Pc--Ne$_N$ the 1D absorption spectrum (Fig.\,\ref{fig2}b) reveals a much broader line profile reminiscent of systems with multiple vibronic transitions.  
Fig.\,\ref{fig3}c shows the respective 2D spectrum of H$_2$Pc--Ne$_N$. 
Here, we observe a single, strongly elongated diagonal peak. 
At the cryogenic temperatures of the Ne cluster environment, only the lowest vibrational state of H$_2$Pc is populated. 
The absence of cross-peaks thus implies the coupling of the ground state to a single vibronic transition in analogy to the H$_2$Pc--He$_N$. 

The broad linear absorption spectrum can be thus purely attributed to ensemble inhomogeneities. 
A quantitative 2D lineshape analysis yields the inhomogeneous and homogeneous line profiles of the S$_1 \leftarrow$ S$_0$ transition in H$_2$Pc--Ne$_N$. 
To this end, we performed a 2D peak fit by adapting the model from Ref.\,\cite{bell_analytical_2015} (details in the Supporting Information). 
The 1D projections of the fit result are shown in Fig.\,\ref{fig3}d, e. 
We find excellent agreement between the fit model and the experimental data, except for an outlier at 15040\,cm$^{-1}$. 
This outlier is not observed in our other data (Fig.\,\ref{fig2}, \ref{fig5}) and is attributed to an experimental artifact. 

We deduce for the inhomogeneous broadening a value of 23\,cm$^{-1}$ and a remarkably narrow homogeneous broadening of $0.42\pm 0.01$\,cm$^{-1}$ (both full-width at half maximum values). 
These values are not limited by the resolution of the experimental apparatus. 
To the best of our knowledge, this is the first experimental determination of the homogeneous broadening in such hetero nanosystems. 
We experimentally determined the fluorescence lifetime to be $> 10\,$ns (not shown, details in Methods section), in good agreement with fluorescence lifetimes in bulk rare gas matrices\,\cite{murray_visible_2011}. 
This implies a negligible lifetime contribution to the homogeneous linewidth. 
The homogeneous broadening thus reflects in good approximation the pure dephasing rate of the system\,\cite{skinner_pure_1986}, which, in analogy to studies in bulk matrices, is attributed to an elastic scattering with phonon modes of the rare gas cluster\,\cite{skinner_pure_1986, friedrich_photochemical_1984}. 
A sample-temperature dependent study of the homogeneous linewidth would provide further insight into the phonon scattering as commonly done in the condensed phase\,\cite{friedrich_photochemical_1984}. 
However, the infrared inactivity and rapid evaporative cooling of the nanoclusters prevent any means for heat injection. 

\begin{figure}
\centering
  \includegraphics{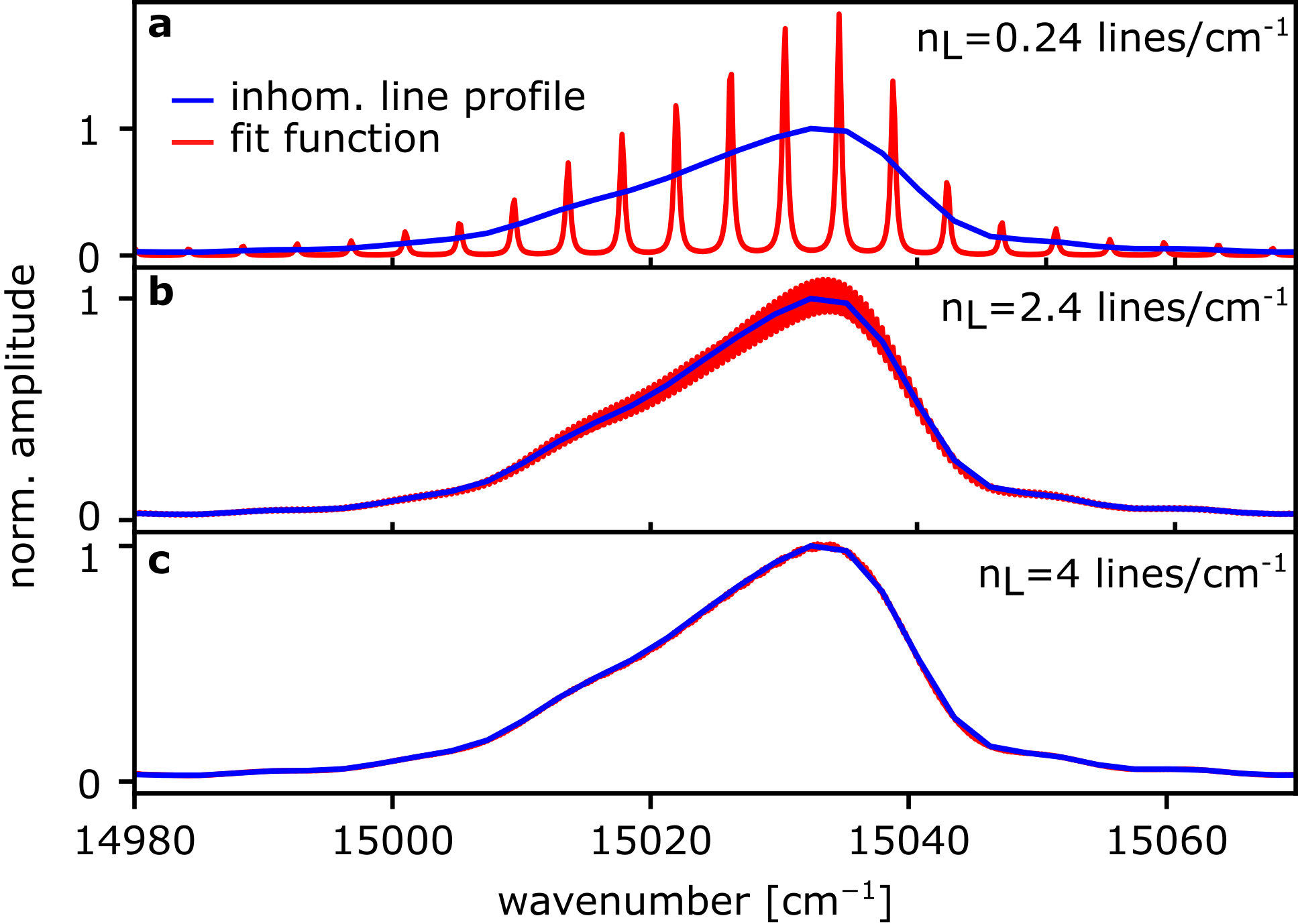}
  \caption{Fit of the inhomogeneous line profile (blue) with a simplistic model of equidistantly spaced homogeneous lines profiles (red). Experimental data (blue), fit result (red). By increasing the line density $n_L$ from top to bottom, the fit converges. For the experimental data, a 1D coherence scan was used, featuring a lower noise level then the 2DES data.}
  \label{fig4}
\end{figure}
The high-resolution data separating the homogeneous and inhomogeneous linewidth allows us to estimate a lower limit for the number of binding configurations between the H$_2$Pc molecule and the cluster surface. 
To this end, we fitted the inhomogeneous lineshape with a simplistic model assuming an equidistant spread of homogeneous line profiles. 
Fit parameters were the density of homogeneous absorption lines $n_L$ and their amplitudes $A_i$. 
We optimized the fit by gradually increasing the line density and considered as convergence criterion the point where the spectral modulations in the fit function are on the level of the noise of the experimental data (Fig.\,\ref{fig4}). 
We find that a minimum line density of $n_L \geq 4.0$\,lines/cm$^{-1}$ (corresponds to 360 configurations) is necessary to fit the data. 
Taking the finite spectral resolution of the experiment into account, this constrains the number of binding configurations to a lower limit of $\geq 216$ (details in the Supporting Information), which corresponds to a mean energetic separation of the binding configurations of only 0.42\,cm$^{-1}$. 

This information about binding configurations is far beyond the resolution and accuracy of current theory and may help to gauge new models of molecule-surface binding configurations, as they play an important role in, e.g. surface chemistry. 
Here, we attribute the large number of binding configurations to the different geometric orientations of the molecule with the cluster lattice and to cluster surface defects (icosahedral cluster structure\,\cite{farges_structure_1981}), while only a minor effect is expected from the statistical cluster size variation in the probed ensemble. 
The latter is rationalized by the fact, that the current study employed fairly small clusters ($\approx 400$ Ne atoms per cluster\,\cite{hagena_condensation_1987}), while basically the same inhomogeneous line profile was obtained for much larger cluster sizes (see nanosecond laser excitation experiment in Fig.\,1 in the Supplementary Information). 
In contrast, experiments in bulk Ne matrices showed an almost one order of magnitude broader inhomogeneous linewidth of $\approx 200\,$cm$^{-1}$\,\cite{murray_visible_2011}. 
The same behavior is observed for 3, 4, 9, 10-perylenetetracarboxylic dianhydride (PTCDA)\,\cite{dvorak_spectroscopy_2012}.  
The much broader inhomogeneous linewidth in the bulk environment may be explained by the formation of a solvation shell when the molecule is fully embedded in the bulk matrix compared to when bound to the surface of a cluster. 
At the same time, the homogeneous broadening of H$_2$Pc-Ne$_\mathrm{N}$ seems in reasonable agreement with the values obtained in bulk matrices of heavier rare-gases where a factor of $\approx 5$ smaller homogeneous broadening is observed\,\cite{geissinger_hole-burning_1992}. 
The lower matrix temperature (4\,K) in these experiments may explain the reduced phonon broadening than observed in our cluster samples with temperatures of $\approx 10$\,K.

2DES also offers time-resolved information about binding configurations (cf.\ Fig.\,\ref{fig1}d). 
The principle is well established in the condensed phase, where thermal fluctuations in the surrounding bath lead to a rearrangement of the binding configurations over time\,\cite{moca_two-dimensional_2015, bizimana_conformational_2017}. 
The spectral diffusion of 2D lineshapes thus mainly reflects the solvent dynamics. 
In contrast, in the cryogenic cluster environment thermal fluctuations are absent and changes in the binding configurations may be directly attributed to the dynamics of the dopant molecules. 
In this context the question arises, if an electronic excitation of the dopant can change the binding configuration with the environment. 
Such mechanisms are common for species dissolved in He droplets, leading e.g. to a migration towards the droplet surface followed by the ejection of the dopant on a pico to femtosecond time scale\,\cite{thaler_femtosecond_2018, thaler_long-lived_2020}. 

In the 2D spectra of H$_2$Pc-Ne$_\mathrm{N}$, the onset of any configuration change would be observable by a broadening of the anti-diagonal linewidth or a reduction of the correlation between excitation and detection frequencies with increasing time $T$. 
We studied the time evolution of the hetero-nanosystem for an evolution time of 0--500\,ps at a reduced spectral resolution (2.3\,cm$^{-1}$ for anti-diagonal linewidth) to limit the measurement time (Fig.\,\ref{fig5}). 
No significant temporal changes can be observed, neither on the correlation coefficient (Fig.\,\ref{fig5}c) nor on the anti-diagonal linewidth (not shown), implying that fluctuations in the binding configurations are below the experimental resolution of 2.3\,cm$^{-1}$ (0.3\,meV). 
This accounts for molecules populated in the electronic ground and excited state, since the response of both is reflected in the 2D spectra. 
We conclude that the molecules are mainly immobile on the cluster surface on the time scale of 500\,ps, which was previously only indirectly confirmed for ground state molecules\,\cite{dvorak_spectroscopy_2012}. 
We note, that the rotation of the cluster in the laboratory frame adds in principle another contribution to the inhomogeneous broadening due to the cluster size distribution in the probed ensemble. 
This broadening is however expected to be negligible due to the large moment of inertia of the clusters. 
\begin{figure}
\centering
  \includegraphics{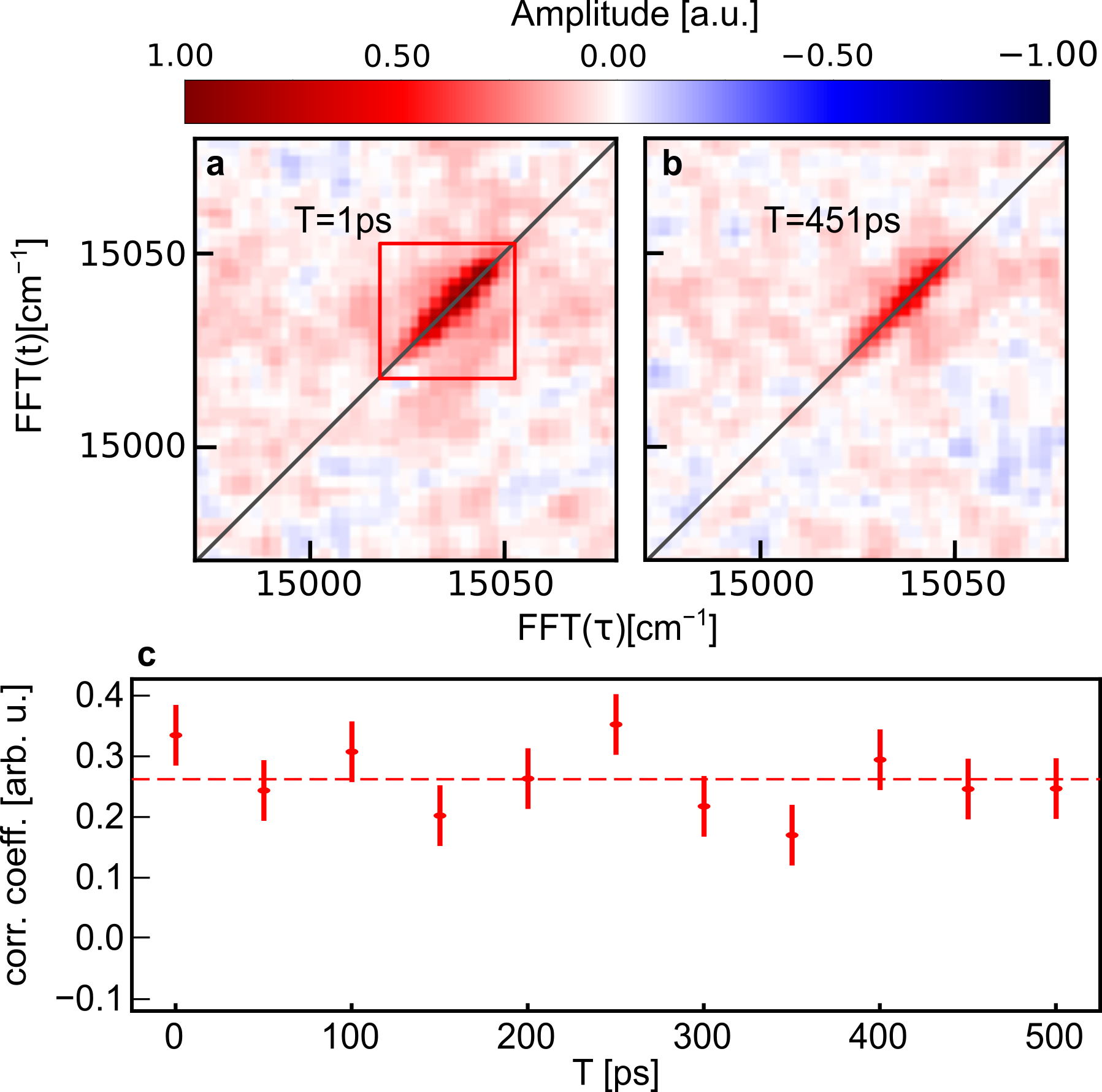}
  \caption{Time evolution of H$_2$Pc--Ne$_N$. 
  (a, b) 2D spectra at evolution times $T=1\,$ps and 451\,ps, recorded at reduced spectral resolution. 
  (c) Correlation coefficient calculated for the marked area in (a) as a function of the evolution time $T$. 
  The dashed line shows the mean correlation value. 
  %Fluctuations in the molecule-surface binding configurations would lead a loss of correlation between the excitation (x-axis) and detection (y-axis) frequencies, which is not observable within the experimental uncertainties. 
}
  \label{fig5}
\end{figure}

On the contrary, dopant species are highly mobile on the surface of superfluid He droplets as reflected in the formation of alkali complexes on the droplet surface\,\cite{schulz_formation_2004}. 
Likewise, species fully immersed in the quantum fluid clusters exhibit high mobility. 
Their rotational motion is well studied and served as a probe of the internal temperature of He clusters\,\cite{hartmann_rotationally_1995}. 
The solvation mechanisms of organic chromophores in superfluid He droplets have been previously studied with high-resolution frequency-domain spectroscopy\,\cite{premke_microsolvation_2014}. 
It was rationalized that solutes are embedded in a non-superfluid He solvation layer. 
The different configuration of the solvation complexes lead to spectral finestructures\,\cite{hartmann_hole-burning_2001, lehnig_fine_2004} while the dispersive solute-cluster interaction causes asymmetric lineshapes due to the statistical cluster size distribution\,\cite{slenczka_inhomogeneous_2001}. 
The homogeneous linewidth for the ZPL of solutes in superfluid He has not been determined to date. 
High-resolution 2DES may solve this problem and provide new insight into the solvation mechanisms in superfluid media. 
For Pc-He$_N$ an inhomogeneous linewidth of $\geq 0.1$\,cm$^{-1}$ was found and an estimate for the homogeneous broadening of 0.02\,cm$^{-1}$ was indirectly deduced\,\cite{slenczka_inhomogeneous_2001}. 
Both are beyond our current experimental resolution. 
This could be solved by deploying advanced sampling strategies\,\cite{sanders_compressed_2012, hutson_ultrafast_2018, agathangelou_phase-modulated_2021} or implementing frequency-comb technology\,\cite{lomsadze_frequency_2017} in order to increase the resolution to the required regime, thus, opening a promising perspective for the study of solvation mechanisms in superfluid environments. 

\section{Conclusions}
The current study explored the nonlinear response of isolated chromphores in different nanosystems with femtosecond time resolution. 
The unique combination of 2DES with cluster-isolation techniques uncovered the homogeneous lineprofile beyond the ensemble average, providing insight into the molecule-surface binding configurations with very high resolution. %, in particular compared to current theory models. 
Other methods for homogeneous linewidth retrieval cannot provide the combined time-frequency resolution of 2DES. 
The presented high-resolution 2DES method, thus, provides a promising perspective for the analysis of molecular dynamics as a function of the local environment with femtosecond time resolution. 
Conversely, the here studied nanosystems represent highly-resolved, well-defined molecular two-level systems which offer ideal conditions to explore many-body mechanisms of high interest in photovoltaics, such as molecular exciton migration and annihilation processes, as well as in quantum information science which is, so far, predominately limited to atomic samples. 
The potential of studying molecular networks on nanoconfined cluster surfaces was recently demonstrated\,\cite{muller_cooperative_2015, izadnia_singlet_2017} and the advantage of 2DES in uncovering intra-particle interactions and cooperative dynamics was proven in previous studies\,\cite{stone_two-quantum_2009, yu_observation_2019, bakulin_real-time_2016}. 

\section{Methods}

\subsection{Sample preparation}
The experimental apparatus including the cluster beam preparation are described in detail in a recent review article\,\cite{bruder_coherent_2019}. 
Further information about the sample preparation is also given in Refs.\,\cite{hagena_cluster_1972, stienkemeier_spectroscopy_2006}.
Briefly, a continuous jet of the rare gas is expanded into vacuum with a stagnation pressure of 50\,bar through a cooled nozzle (5\,$\mu$m orifice).
The nozzle temperature is 14\,K for He (mean cluster size of N$_\text{He}$=15000 atoms\,\cite{toennies_superfluid_2004}) and 70\,K for Ne (N$_\text{Ne}$=400\,\cite{hagena_condensation_1987}), respectively. 
The cluster beam is skimmed by a 400\,$\mu$m conical skimmer positioned in a distance of 15\,mm from the nozzle. 
The cluster beam is doped in the adjacent chamber, where it passes a 10\,mm-long, heated oven cell containing H$_2$Pc powder (Sigma Aldrich, 29H,31H-Phthalocyanine, 98\%). 
The doping follows a statistical pick-up process described by Poissonian statistics and is controlled by the cell temperature and, thus, the density of evaporated molecules. 
The cell temperature is optimized for maximum fluorescence signal, which corresponds to molecule densities of 5\,$\times10^{12}$\,cm$^{-3}$ (350\,$^{\circ}$C cell temperature) for He and 3.2\,$\times10^{13}$\,cm$^{-3}$ (380\,$^{\circ}$C) for Ne, respectively. 
At these conditions, the estimated mean dopant number per cluster are 1 for He and 2-3 for Ne clusters.
Evaporation of cluster atoms cools the nanosystems afterwards to equilibrium temperatures of 0.37\,K for He\,\cite{hartmann_rotationally_1995}, and $\approx 10$\,K for Ne\,\cite{farges_structure_1981}. 
The base pressure in the doping chamber was $1.48 \times 10^{-6}\,$mbar which is due to the effusively emitted hot molecules from the oven cell. 
The doped cluster beam passes an orifice of 5\,mm into the next chamber to suppress the latter contribution. 
In this chamber ($3.8\times 10^{-8}$\,mbar base pressure) the femtosecond laser pulses intersect with the cluster beam at right angle and the  sample fluorescence is imaged with a lense onto a photo multiplier tube mounted in perpendicular direction to the laser and cluster beam propagation. 
The detector arrangement collects $\approx 17\%$ of the total fluorescence. 
The detector response time is 0.57\,ns. 
The fluorescence lifetime was determined by tracking the decay of the fluorescence yield with a fast digital-to-analog converter (bandwidth: 500\,MHz, sampling rate 1\,Gs/s). 

\subsection{Spectroscopy method}
The high-repetition-rate femtosecond laser system and phase-modulated 2DES setup is described in detail elsewhere\,\cite{bruder_coherent_2019, uhl_coherent_2021}. 
Laser parameters were: center wavelength 660--670\,nm, spectral width 25\,nm (FWHM), pulse duration $\approx 50$\,fs %(estimated from SHG-FROG in front of the vacuum chamber and adding the additional dispersion of the vacuum window)
, pulse energies 25--35\,nJ per pulse, laser focus diameter 200\,$\mu$m (both at the target) and laser repetition rate 200\,kHz. 
A typical pulse spectrum is shown in Fig.\,\ref{fig2}b. 

For the 2DES experiments, the cluster beam was excited by a collinear pulse train of 4 femtosecond laser pulses and the fluorescence is detected as parametric function of the inter-pulse delays $\tau, T, t$. 
A 2D Fourier transform with respect to $\tau, t$ yields the rephasing and non-rephasing 2D spectra of which the absorptive correlation spectrum is computed and shown in the main text, except for Fig.\,\ref{fig3}c showing the real-part of the rephasing spectrum. 
For the 1D coherence scans pulses 3,4 were blocked and the real-part of the 1D Fourier transform with respect to $\tau$ is shown in the main text. 
For the 1D coherence scan in Fig.\,\ref{fig2}, $\tau$ was scanned from 0 to 120 ps in 100\,fs steps.
For the 2DES measurements the coherence times $\tau, t$ were scanned, from 0 to 50\,ps in steps of 500\,fs (Fig.\,\ref{fig3}a,b), from 0 to 42\,ps in steps of 300\,fs (Fig.\,\ref{fig3}c) and from 0 to 6\,ps in steps of 300\,fs (Fig.\,\ref{fig5}). 
For the measurements in Fig.\,\ref{fig5} the population time $T$ was scanned from 1 to 501\,ps. 
Values $T<100$\,fs were omitted to avoid the influence of parasitic pulse overlap effects.

To increase the signal-to-noise ratio, the carrier-envelope phase of each pulse is modulated at frequencies $\Omega_i$ ($i=1-4) \approx$ 110\,MHz on a shot-to-shot level using acousto-optical modulators. 
This leads to well-defined beat notes in the fluorescence signals of $\Omega_{21}=5$\,kHz, $\Omega_{43}=8$\,kHz for the linear excitation (1D coherence scan) and $\Omega_{21}\pm\Omega_{43}=13$\,kHz for the fourth-order non-rephasing and  3\,kHz  for the rephasing signal, respectively ($\Omega_{ij}$ denotes $\Omega_i-\Omega_j$). 
The beat signals are efficiently filtered and amplified by lock-in detection. 

As reference signal for the lock-in detection, a portion of the pulse train is spectrally filtered in a monochromator (0.025\,nm spectral resolution) at a frequency of $\nu_R=c/662.33$\,nm and detected with a PMT. 
The reference signal records the optical interference of the pulses and, thus, tracks the phase changes and phase jitter in the optical setup at the frequency $\nu_R$. 
The heterodyne demodulation of the fluorescence signal $S(\tau, T, t, \Omega_{21}\pm\Omega_{43})$ with the reference signal $R(\tau, T, t, \Omega_{21}\pm\Omega_{43})$ inside the lock-in amplifier removes the modulation as well as the phase jitter from the signal $S$, which provides the required interferometric stability for the measurements. 
In addition, $S$ is detected in the rotating frame of $R$, which down-shifts the signal frequencies $\nu_S$ by $2$ orders of magnitude to $\bar{\nu}=\nu_S-\nu_R$. 
To recover the absolute frequencies in the Fourier spectra, the frequency axis are up-shifted again by $\nu_R$. 

For the given monochromator resolution, the reference interference signal is detectable for $\tau, t$ delays up to 120\,ps, which determines the attainable Fourier resolution in the experiment. 
To avoid Fourier-transform artifacts, the 1D coherence scan data of the H$_2$Pc--He$_N$ sample was apodised in the time domain with an exponential function decaying to 5\% of its amplitude at the end of the delay scan range. 
The 2DES data of the H$_2$Pc--He$_N$ sample was apodised by a 2D Gaussian decaying to 5\% of its amplitude along each dimension. 
The apodisation leads to an artificial broadening of the spectral lines reflecting the limited resolution of the experimental apparatus. 
Accordingly, for the 1D coherence scan, the experimental resolution was 0.3\,cm$^{-1}$ and for the 2DES measurements (reduced scan range) it was 0.65\,cm$^{-1}$ (Fig.\,\ref{fig4}) and 4.6\,cm$^{-1}$ (Fig.\,\ref{fig5}) in each Fourier dimension. 
Note, that the resolution along the diagonal (anti-diagonal) dimension is higher\,\cite{bell_analytical_2015}. 
In case of the H$_2$Pc--Ne$_N$ sample, data apodisation was not necessary and the inhomogeneous (homogeneous) linewidths of the sample were determined without spectral broadening introduced by the experimental apparatus. 

\bibliography{PcH2-Paper_UB,H2Pc_2D_paperLB}% Produces the bibliography via BibTeX.

\section*{Data availability}
The data that support the findings of this study are available from the corresponding author upon reasonable request.

\section*{Acknowledgements}
Funding by the European Research Council within the Advanced
Grant of " COCONIS " ( 694965 ) and by the Deutsche
Forschungsgemeinschaft ( IRTG 2079 ) are acknowledged. 
We thank M. Walter and S. Ferchane for useful discussions about DFT calculations of the Ne--H$_2$Pc interaction.

\section*{Competing interests}
The authors declare no competing interests. 

\section*{Additional information}
Supplementary information is available for this paper. 

\end{document}

\title{Supplementary information: \\High-resolution two-dimensional electronic spectroscopy reveals homogeneous line profiles in isolated nanoparticles}

\author{Ulrich Bangert}
\author{Frank Stienkemeier}
\author{Lukas Bruder\thanks{lukas.bruder@physik.uni-freiburg.de}}%

\affil{Institute of Physics, University of Freiburg, Hermann-Herder-Str. 3, 79104 Freiburg, Germany.}
\date{}

\maketitle

\section{1D coherence scans}
\begin{figure}
\centering
  \includegraphics[width=\textwidth]{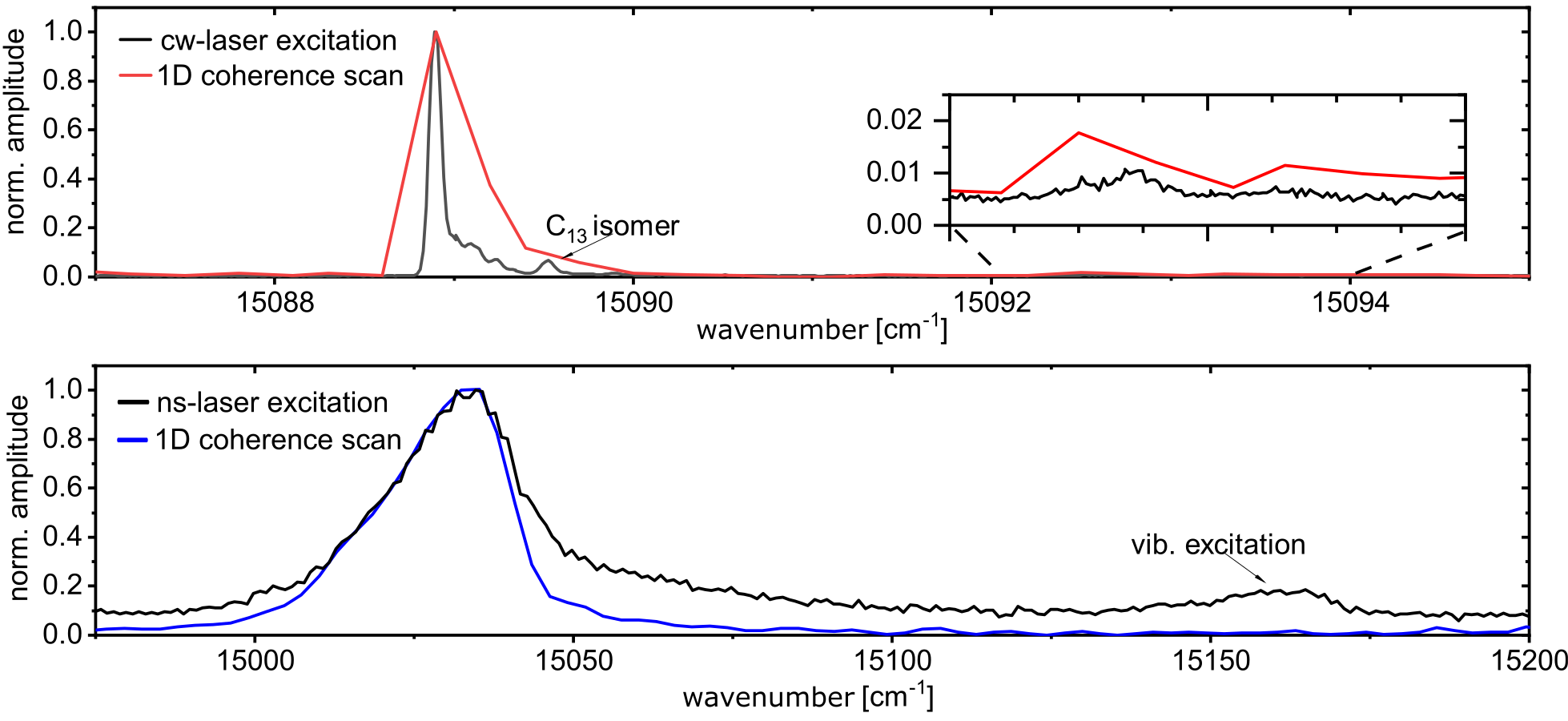}
  \caption{Fourier spectra of 1D coherence scans of H$_2$Pc--He$_N$ (red) and H$_2$Pc--Ne$_N$ (blue) compared to LIF excitation spectra using narrow-band cw and nanosecond lasers (black). 
  The inset shows the phonon wing. 
  The H$_2$Pc--He$_N$ reference spectrum was supplied by A. Slenczka\,\cite{fischer_unpublished_2020}. 
  The one of H$_2$Pc--Ne$_N$ was recorded in our laboratory with the experimental setup described in Ref.\,\cite{dvorak_spectroscopy_2012} at an oven cell temperature of 390\,$^{\circ}$C and laser pulse energy of 109\,$\mu$J. 
The high pulse energy in the latter experiment leads to saturation broadening of the strong absorption lines and a relative enhancement of the weak vibronic lines at 15160\,cm$^{-1}$.}
  \label{fgr:sfig1}
\end{figure}

Fig.\,\ref{fgr:sfig1} shows a zoom of the Fourier spectra from the 1D coherence scans of H$_2$Pc--He$_N$ and H$_2$Pc--Ne$_N$ (cf. Fig.\,2b in the main text) along with excitation spectra obtained by light-induced fluorescence (LIF) measurements using continuous wave (cw) and nanosecond lasers. 
These spectra feature very narrow excitation bands and a distinct blue-shift compared to H$_2$Pc spectra in solution (Fig.\,2b in the main text) or of hot H$_2$Pc molecules in the gas phase\,\cite{eastwood_spectra_1966}. 
Due to the low internal temperature of the molecules in the cluster-isolation experiments only the vibrational ground state of H$_2$Pc is populated,  which strongly curtails the number of possible vibronic transitions and reduces the excitation spectrum mainly to the ZPL of the Q$_x$ absorption band. 
Due to the weakly perturbing cluster environment, the ZPL is hardly red shifted ($\leq 100$\,cm$^{-1}$) with respect to the spectrum of cold gas phase molecules: ZPL in H$_2$Pc--He$_N$ is at 15088.9\,cm$^{-1}$, in H$_2$Pc--Ne$_N$ at 15032\,cm$^{-1}$ in in jet-cooled H$_2$Pc at 15132\,cm$^{-1}$\,\cite{fitch_fluorescence_1980}, hence, explaining the strong blue shift compared to experiments in solution.  

In the H$_2$Pc--He$_N$ data, the ZPL features a spectral wing on the high-frequency side, which can be identified as a C$_{13}$-isotope of H$_2$Pc (ZPL at 15089.5\,cm$^{-1}$)\,\cite{lehnig_fine_2004}. 
In the H$_2$Pc--Ne$_N$ spectrum this feature is obscured by the inhomogeneous broadening. 
Moreover, the phonon wing, a spectral feature attributed to the excitation of volume vibrations of the helium environment, appears weakly at approximately 15093\,cm$^{-1}$.
However, its relative amplitude of 1\,\%, known from the cw laser LIF spectrum, is close to the SNR of the 1D coherence scan.
Other resonances are not observed in the H$_2$Pc--He$_N$ spectra. 
Only at two orders of magnitude higher laser intensities (not shown) strongly saturating the ZPL, a weak coupling to vibrational modes of the molecule has been observed in the steady-state LIF measurements~\cite{lehnig_fine_2004}. 
Likewise, in H$_2$Pc--Ne$_N$ a weak vibronic coupling is only observed in the nanosecond laser experiment, where the optical transition was strongly saturated to enhance weak features. 

\section{2D lineshape fit}
\begin{figure}
\centering
  \includegraphics{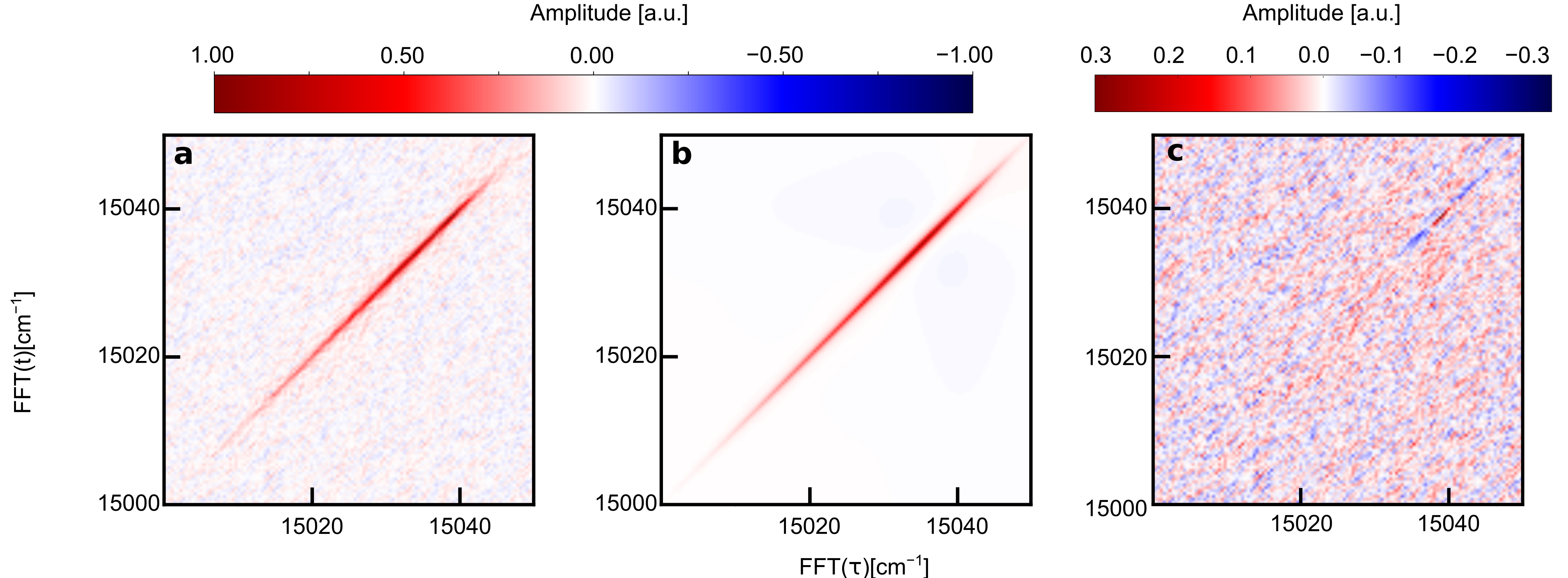}
  \caption{2D peak fit. (a) experimental H$_2$Pc--Ne$_N$ data, (b) fitted function, (c) residues between (a) and (b).}
  \label{fgr:sfig2}
\end{figure}

In 2DES, the spectral line profile along the diagonal(anti-diagonal) correspond to the inhomogeneous(homogeneous) absorption profiles of the system\,\cite{jonas_two-dimensional_2003}. 
To extract both linewidths, we performed a 2D peak fit of the H$_2$Pc--Ne$_N$ spectrum from Fig.\,3c in the main text. 
To this end, we adapted the 2D fit function calculated by Bell et al.\,\cite{bell_analytical_2015} and fitted the rephasing part of the 2D spectrum.
In the homogeneously broadened dimension the fit function consists of a single Lorentzian profile, in the inhomogeneously broadened dimension we apply a sum of two Gaussian functions to account for the double peak structure (main peak and red shoulder) of the inhomogeneous profile, leading to the 2D line shape function:
\begin{align}
    \begin{split}
      S_R(\omega_t,\omega_\tau) = \sum_{i}^{1,2}& A_i \frac{1}{2\sigma_i(2\gamma-i(\omega_t+\omega_\tau))}\\
        & \times [e^{\frac{(\gamma-i(\omega_t-\omega_{0i}))^2}{2\sigma_i^2}}\text{Erfc}(\frac{\gamma-i(\omega_t-\omega_{0i})}{\sqrt{2}\sigma_i})\\
        & +e^{\frac{(\gamma-i(-\omega_\tau+\omega_{0i}))^2}{2\sigma_i^2}}\text{Erfc}(\frac{\gamma-i(-\omega_\tau+\omega_{0i})}{\sqrt{2}\sigma_i})]\, .
        \label{eq_fit}
    \end{split}
\end{align}
Here $A_i$, $\omega_{0i}$ and $\sigma_{0i}$ are the amplitudes, center frequencies and widths of the Gaussians.
$\gamma$ is the dephasing rate which is connected to the homogeneous line width by FWHM\,=\,2$\gamma$.
The 2D fit result is shown in Fig.\,\ref{fgr:sfig2} along with the residues.
Except for an outlier in the experimental data at 15040\,cm$^{-1}$, we get an excellent agreement between the experimental data and the fit result as indicated by the residues. 
%Moreover, fitting the 1D spectrum (Fig.\,\ref{fig:sfig1}) with the same double-Gaussian function, we obtain the same fit result for the inhomogeneous line profile (not shown), confirming the irrelevance of the experimental outlier in the 2D spectrum for the fit procedure. 
The homogeneous line width is deduced directly from the 2D fit. 
The inhomogeneous line width is evaluated from the FWHM value of the 1D coherence scan. 

\section{Estimating the minimum number of binding configurations}
To evaluate the number of binding configurations in the H$_2$Pc--Ne$_N$ system, we fitted the inhomogeneous line profile of the 1D coherence scan (Fig.\,\ref{fgr:sfig1}b) with the simplistic fit model described in the main text. 
For the fit, we consider the spectral interval 14980-15070cm$^{-1}$, where the H$_2$Pc--Ne$_N$ spectrum shows a significant amplitude. 
The fit yields a minimum density of homogeneous absorption lines of $\geq 4$\,lines$/$cm$^{-1}$, amounting to a total number of 360 binding configurations. 
As the transition frequencies of the individual binding configurations are not likely distributed equidistantly, an even higher line density and, consequently, a higher number of configurations should be taken into account. 
On the other hand, the finite spectral resolution of the experiment introduces an uncertainty towards lower numbers of binding configurations. 
Since we measured the line profile in the time domain over a finite time window (3\,ps), a small uncertainty remains that the line profile exhibits a spectral substructure which is neither resolved in our experiment nor in the previous nanosecond laser LIF experiments. 
A sub-structure would lead to a lower line density required for the fitting of the data. 
Taking this factor into account, we compute an estimate for the lower limit of the number of binding configurations of $\geq 216$, which corresponds to a mean energetic separation of the binding configurations of 0.42\,cm$^{-1}$. 
In principle, the C$_{13}$-isotopes of H$_2$Pc should be also taken into account for the fit procedure. 
However, their small spectral amplitudes ($\leq 10\%$ of ZPL) should make a negligible contribution and were therefore not considered. 

Intuitively, the high number of configurations could be explained with the intrinsic properties of the neon clusters and the doping process. 
The broad distribution of cluster sizes and the icosahedral structure of the clusters form a large variety of slightly different surfaces the H$_2$Pc can attach to. 
In the nanosecond laser experiments (Fig.\,\ref{fgr:sfig1}b) much larger clusters were used and essentially the same inhomogeneous line profile has been obtained. 
Likewise, in similar studies using PTCDA molecules no significant effect of the cluster size on the inhomogeneous line profile was found\,\cite{dvorak_spectroscopy_2012}. 
From this, we conclude, that the statistical cluster size distribution has a minor impact on the molecule-surface binding configurations and surface defects of the icosahedral geometry play a larger role. 
Further, each surface can have local potential minima depending on the orientation and position of the molecule.
The molecules are likely trapped in these local minima, due to the random pick-up process and the rapid evaporative cooling afterwards.

%\nocite{*}
\bibliography{H2Pc_2D_paperLB_SI}